\documentclass[twocolumn,apj,iop]{openjournal}

\DeclareRobustCommand{\VAN}[3]{#2}
\let\VANthebibliography\thebibliography
\def\thebibliography{\DeclareRobustCommand{\VAN}[3]{##3}\VANthebibliography}

\usepackage{graphicx}	

\usepackage{orcidlink}

\begin{document}

\title{Hydrogen-Atmosphere White Dwarfs Are Less Likely To Be Found with Wide-Binary Companions}
\shorttitle{WD Binaries}
\author{\vspace{-0.5in}Jeremy Heyl\orcidlink{0000-0001-9739-367X}$^1$}
\shortauthors{Heyl}

\affiliation{$^1$University of British Columbia, Vancouver, BC, Canada}

\begin{abstract}
The fraction of white dwarfs found in wide binaries is estimated by cross referencing a catalogue of wide binaries with catalogues photometrically determined white-dwarf candidates and spectroscopically confirmed white dwarf stars.  The wide-binary fraction of white dwarfs with hydrogen-dominated atmospheres is about 5\%, but the fraction of white dwarfs with helium or carbon-dominated atmospheres is significantly larger.   Using spectroscopic classifications, the binary fraction of DA white dwarfs is determined to be $0.063\pm0.002$ and for non-DA white dwarfs is larger at $0.080\pm0.004$.
\end{abstract}


\maketitle


\section{Introduction}
\label{sec:intro}

Most stars are members of multiple stellar systems, and most stars will become white dwarfs when they exhaust their nuclear fuel, so it is quite natural to examine the fraction of white dwarfs in multiple stellar systems.  Additionally the bulk of white dwarfs have substantial hydrogen atmospheres while a smaller fraction have helium or carbon atmospheres.  \citet{2024MNRAS.529.2910N} has examined the correlation of binarity with the metal abundance within white dwarf atmospheres to seek hints of the role of companions in the metal pollution.   However, perhaps a more basic question is: what fraction of white dwarfs are actually in binaries and whether white dwarfs with different atmospheric compositions have different binary fractions?

The \textit{Gaia} satellite \citep{2016A&A...595A...1G} and its successive data releases \citep{2016A&A...595A...2G,2018A&A...616A...1G,2021A&A...649A...1G,2023A&A...674A...1G} have driven a revolution of our understanding of stellar astrophysics and especially of white dwarfs.  \citet[][GF]{2021MNRAS.508.3877G} has developed a catalogue of white dwarf candidates from Gaia EDR3 that provides a tentative classification of the white dwarfs as having hydrogen or helium atmospheres.  Furthermore, \citet{2021MNRAS.506.2269E} created a catalogue of high-probability wide binaries from Gaia EDR3.  The Montreal White Dwarf Database \citep[][MWDDB]{2017ASPC..509....3D} provides a complementary census of the white-dwarf population with many white dwarfs classified by their spectral type which yields more precise determination of the atmospheric composition.  Cross referencing the \citet{2021MNRAS.506.2269E} binary catalogue with these two white-dwarf catalogues yields a catalogue of likely and confirmed white-dwarf stars that are very likely to have companions.  Furthermore, these resultant catalogues yield estimates of the binary fraction of white dwarfs as a function of atmospheric composition, spectral type and mass.

\section{Data}
\label{sec:data}

The \citet{2021MNRAS.506.2269E} binary catalogue applies astrometric and photometric cuts on the entire Gaia EDR3 catalogue before seeking wide binary companions from spatial and kinematic coincidences among the sources in the catalogue.  In particular, the constraints are that the source must have a measured Gaia $G$-band magnitude, its parallax must exceed 1~mas, the error in the parallax must be less than 2~mas, and the ratio of the measured parallax to the error in the parallax must exceed five.  
To estimate the binary fraction of the white dwarfs, these constrains are also applied to the GF and MWDDB catalogues.  Finally, both the  GF catalogue and the MWDDB are cross referenced with the primary and secondary stars listed in the \citet{2021MNRAS.506.2269E} catalogue using the Gaia EDR3 source id that all three catalogues use in common to yielding a catalogue of candidate wide binaries containing white-dwarf stars.

\section{Results}
\label{sec:results}

The two white dwarf catalogues trade completeness against accuracy.  The GF catalogue yields a probability that a particular Gaia EDR3 source is a white dwarf and estimates of the mass and atmospheric composition obtained from fitting photometric data to model spectral energy distributions.  The portion of the MWDD used in this study gives spectral classifications of the white dwarfs and mass estimates obtained by fitting spectroscopic data to model spectra. 

\subsection{Photometric Classification}
\label{sec:photo}

From the GF catalogue, a sample of probable white-dwarf stars is obtained by selecting only those sources where the white-dwarf probability, $P_\textrm{\scriptsize WD}$, is greater than 0.99.   From this sample, selecting those sources where the hydrogen model yields a significantly better fit than the helium model yields a subsample of white dwarfs likely to have hydrogen atmospheres. Conversely, a sample of helium atmosphere (or non-hydrogen atmosphere) white dwarfs is built by selecting sources where the helium model yields a significantly better fit than the hydrogen model.   In both cases, objects where a mixed model is preferred are excluded.  Each of these subsamples are cross-referenced against the binary catalogue, and sources with chance coincidences greater than 10\% ($R_\textrm{\scriptsize chance}$) are excluded. Table~\ref{tab:photo} outlines the down-selections to create the photometric catalogues.
\begin{table}
\centering
\caption{Results of Photometric Classification}
\label{tab:photo}
\begin{tabular}{lr}
\hline
\hline
Number of objects in GF main catalogue                                    &  1280266 \\
~~with $P_\textrm{\scriptsize WD}>0.99$                                   &  192613 \\
\hline
\multicolumn{2}{c}{Hydrogen Atmospheres} \\
\hline
~~with $\chi^2_\textrm{\scriptsize H} < \chi^2_\textrm{\scriptsize He}-1$ & 53364 \\
~~meet binary catalogue astrometric criteria                              & 50950 \\
~~with $\chi^2_\textrm{\scriptsize H} < \chi^2_\textrm{mix}$              & 50871 \\
~~with a wide binary companion                                            &  3210 \\
~~with $R_\textrm{\scriptsize chance}<0.1$                                &  2381 \\
~~binary fraction                                                         & $0.0468\pm0.0009$ \\
\hline
\multicolumn{2}{c}{Helium Atmospheres} \\
\hline
~~with $\chi^2_\textrm{\scriptsize He} < \chi^2_\textrm{\scriptsize H}-1$ &  7662 \\
~~meet binary catalogue astrometric criteria                              &  7158 \\
~~with $\chi^2_\textrm{\scriptsize He} < \chi^2_\textrm{mix}$             &  5351 \\
~~with a wide binary companion                                            &   753 \\
~~with $R_\textrm{\scriptsize chance}<0.1$                                &   640 \\
~~binary fraction                                                         & $0.120\pm0.004$ \\
~~difference                                                              & $+21\sigma$ \\
\hline
\end{tabular}
\end{table}

The ratio of white dwarfs whose photometry is best fit by hydrogen models to those best fit by helium models is about eleven to one among the entire sample.  This is much larger than the ratio of DA white dwarfs to non-DAs (i.e. DB, DC, DO and DQ) which is about three or four to one \citep[][and \S~\ref{sec:spect}; c.f. \citealt{2022MNRAS.509.2674G} found a ten-to-one ratio]{2020ApJ...898...84K}; therefore, possibly many of objects classified as hydrogen atmosphere white dwarfs actually do not have hydrogen atmospheres. Looking more closely, the binary fraction of the helium-atmosphere white dwarfs is about 12\% compared to less than 5\% for those white dwarf best fit by hydrogen atmospheres.  Given the sizes of the samples, this difference is significant at the $21-\sigma$ level (the significance is calculated directly from the binomial distribution, not the Poisson or normal approximations).  However, does this indeed imply that white dwarfs with hydrogen atmospheres are nearly three times less likely to be in a wide binary than those with helium atmospheres?

\begin{figure}
    \centering
    \includegraphics[width=\linewidth]{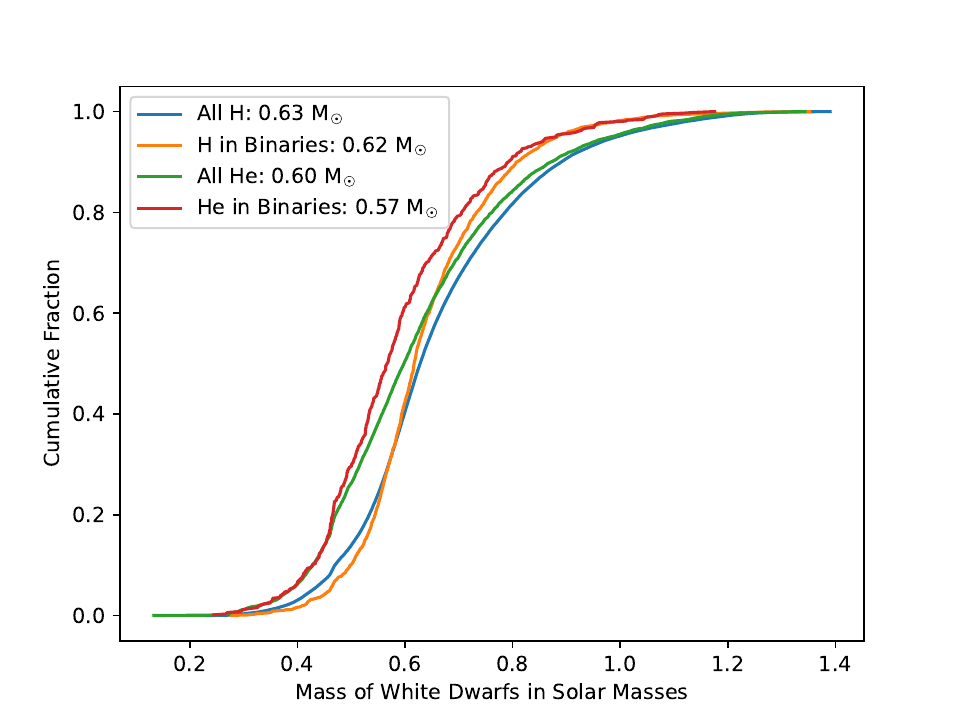}
    \caption{The Cumulative Distribution of Masses of Hydrogen and Helium Atmosphere White Dwarfs Obtained by Photometric Modelling.  The median mass of each population is presented in the legend.}
    \label{fig:cmass_phot}
\end{figure}

\begin{figure}
    \centering
    \includegraphics[width=\linewidth]{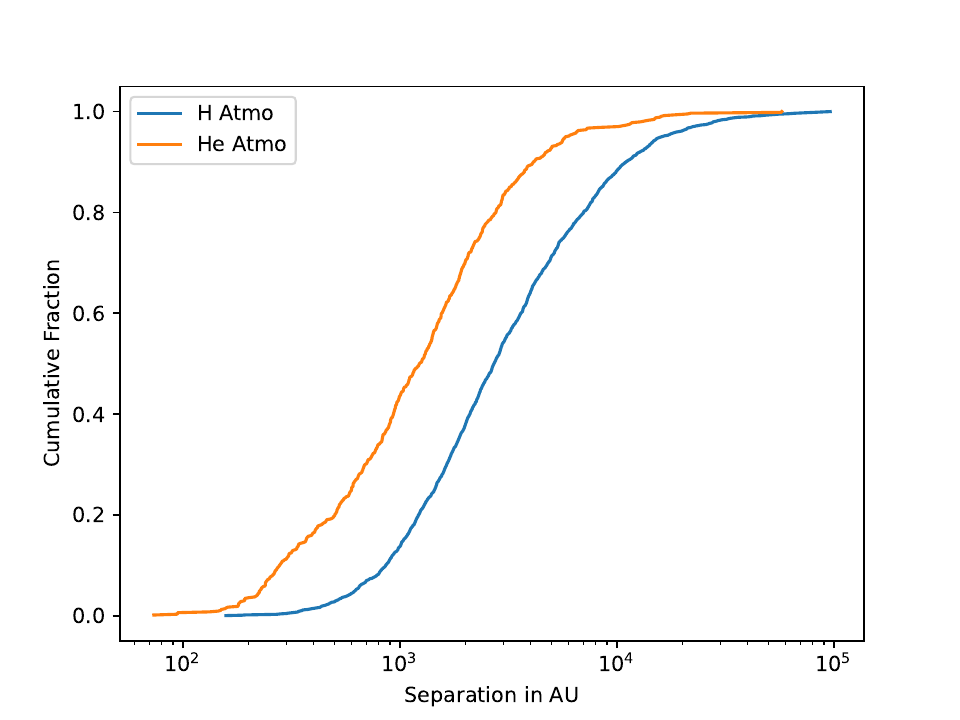}
    \caption{The Cumulative Distribution of Projected Binary Separations of Hydrogen and Helium Atmosphere White Dwarfs Identified by Photometric Modelling}
    \label{fig:sep_photo}
\end{figure}

Figure~\ref{fig:cmass_phot} depicts the cumulative mass distribution (using the masses determined from the photometric fits) of the white dwarfs with best-fitting hydrogen and helium models and within those two classes the objects with wide-binary companions.  The stars best-fitted with hydrogen atmospheres typically have larger masses than those best-fitted with helium atmospheres.  This contradicts what one expects from spectroscopic determinations of the white-dwarf atmospheres where DB white dwarfs typically have larger masses than DA \citep[][and\S~\ref{sec:spect}]{2019ApJ...882..106G,2020ApJ...898...84K}. Furthermore, for both categories, the masses of those stars with companions are typically smaller than those without companions.  A second diagnostic is to examine the distribution of projected separations between the white dwarfs and their companions (Fig.~\ref{fig:sep_photo}).  The companions of those white dwarfs best fitted by helium atmospheres are typically closer than those best-fitted with hydrogen atmospheres.  The median angular distance between a white dwarf fitted with a hydrogen atmosphere is 11 arcseconds, and for a helium atmosphere the corresponding distance is just 3.5 arcseconds.  These results indicate that perhaps the photometric fitting is affected by the presence of a companion: the presence of a companion may increase the likelihood that a star is better fitted by a helium atmosphere and by a lower mass model.  Although the details are somewhat different, these distributions are similar to the distributions for subsamples within 100~pc and for subsamples with $u-$band photometry, which may allow for more precise determination of the atmospheric parameters from photometry; consequently, these trends bear further scrutiny through the examination of a sample of spectroscopically classified white dwarfs.

\subsection{Spectroscopic Classification}
\label{sec:spect}

The MWDDB provides spectroscopic classifications of white dwarf atmospheres as well as estimates of their masses.   The database contains 732 unique spectral types.  Among those white dwarfs that meet the astrometric criteria of the binary catalogue (43,324), more than half (21,963) fall within the following spectral types: DA, DB, DC, DO and DQ.  The catalogue contains subtypes of these classes such as (DA1-2, DA3.0 and DB1.8) which are included in the corresponding major type such as DA, but stars with other non-numerical subtype designations (such as DAH or DAZ) and spectroscopic binaries are excluded from this analysis, resulting in the numbers of white dwarfs of various classes shown in Table~\ref{tab:spect}.  Again for all of the non-DA classes of white dwarfs, the binary fraction is significantly larger 8.0\% compared to 6.3\% for DA stars, a nearly five-sigma difference.  The result of a binary fraction of 6.3\% for DA stars agrees with the result of \citet{2024MNRAS.529.2910N} of $6.8^{+3.1}_{-2.4}\%$ for a sample of 82 stars.  Of course, this is much less dramatic than the result for the photometric classification.
\begin{table}
\centering
\caption{Results of Spectroscopic Classification} 
\label{tab:spect}
\begin{tabular}{lr}
\hline
\hline
Number of objects in MWDDB                                                &   70681 \\
~~meet binary catalogue astrometric criteria                              &   43324 \\
\hline
DA classification                                                         &  17692 \\
~~with a wide binary companion                                            &  1370 \\
~~with $R_\textrm{\scriptsize chance}<0.1$                                &  1109 \\
~~binary fraction                                                         & $0.063\pm0.002$ \\
\hline
DB classification                                                         &  1277 \\
~~with a wide binary companion                                            &   115 \\
~~with $R_\textrm{\scriptsize chance}<0.1$                                &    89 \\
~~binary fraction                                                         & $0.070\pm0.007$\\
~~difference                                                              & $+1.1\sigma$ \\
\hline
DC classification                                                         &  2533 \\
~~with a wide binary companion                                            &   233 \\
~~with $R_\textrm{\scriptsize chance}<0.1$                                &   211 \\
~~binary fraction                                                         & $0.083\pm0.005$\\
~~difference                                                              & $+4.1\sigma$ \\
\hline
DO classification                                                         &    42 \\
~~with a wide binary companion                                            &     3 \\
~~with $R_\textrm{\scriptsize chance}<0.1$                                &     1 \\
~~binary fraction                                                         & $0.02\pm0.02$\\
~~difference                                                              & $-0.7\sigma$ \\
\hline
DQ classification                                                         &   419 \\
~~with a wide binary companion                                            &    42 \\
~~with $R_\textrm{\scriptsize chance}<0.1$                                &    42 \\
~~binary fraction                                                         & $0.10\pm0.01$\\
~~difference                                                              & $+3.0\sigma$ \\
\hline
Subtotal of Non-DA classifications above                                  &  4271 \\
~~with a wide binary companion                                            &   393 \\
~~with $R_\textrm{\scriptsize chance}<0.1$                                &   343 \\
~~binary fraction                                                         & $0.080\pm0.004$\\
~~difference                                                              & $+4.6\sigma$ \\
\hline
\hline
\hline
\end{tabular}
\end{table}

\begin{figure}
    \centering
    \includegraphics[width=\linewidth,clip,trim=0 0.5in 0 0]{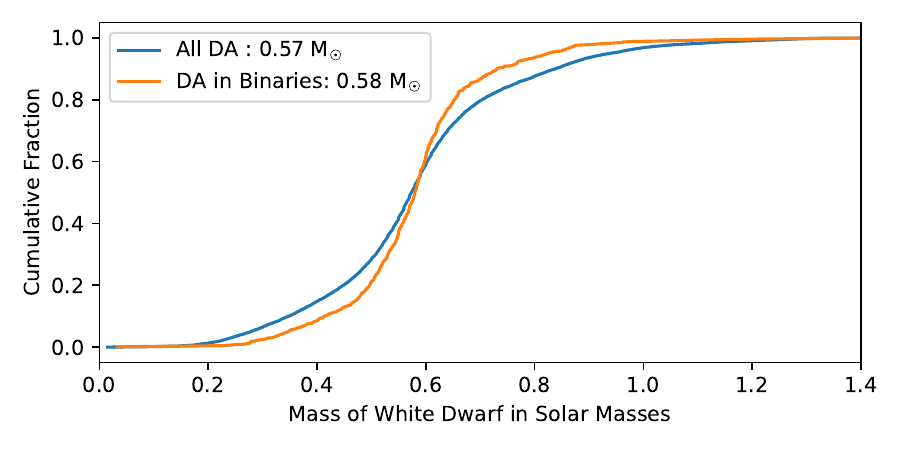}
    \includegraphics[width=\linewidth,clip,trim=0 0.5in 0 0.2in]{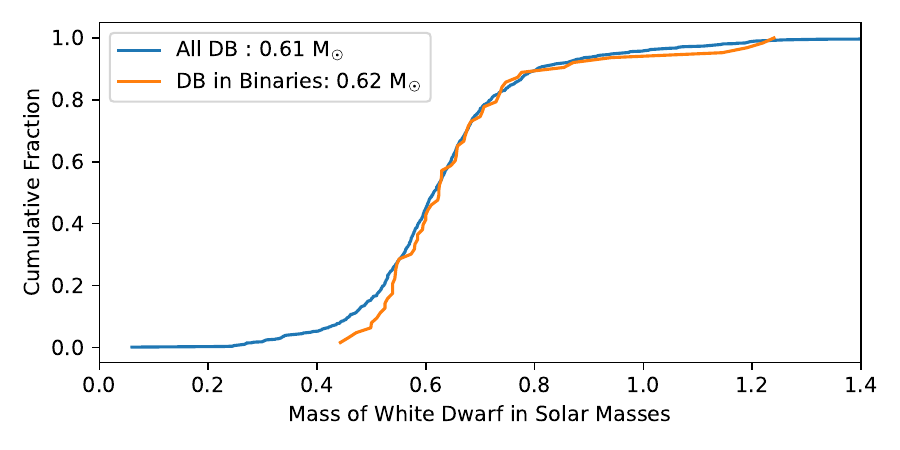}
    \includegraphics[width=\linewidth,clip,trim=0 0.5in 0 0.2in]{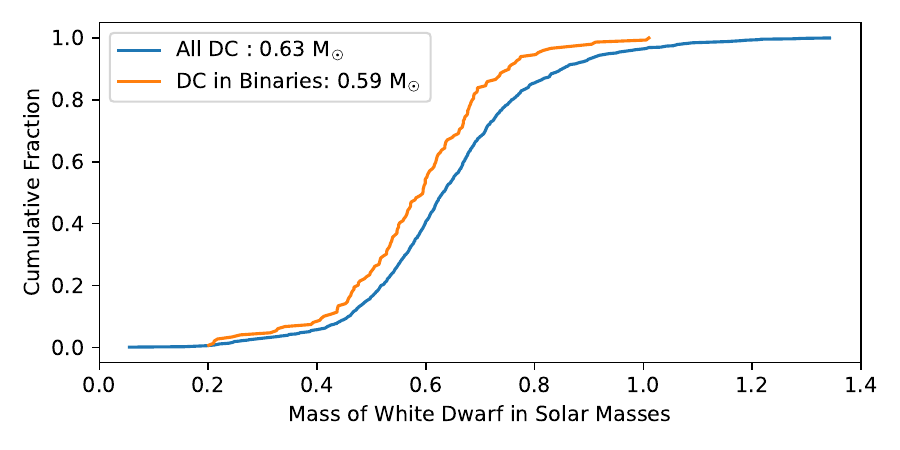}
    \includegraphics[width=\linewidth,clip,trim=0 0.5in 0 0.2in]{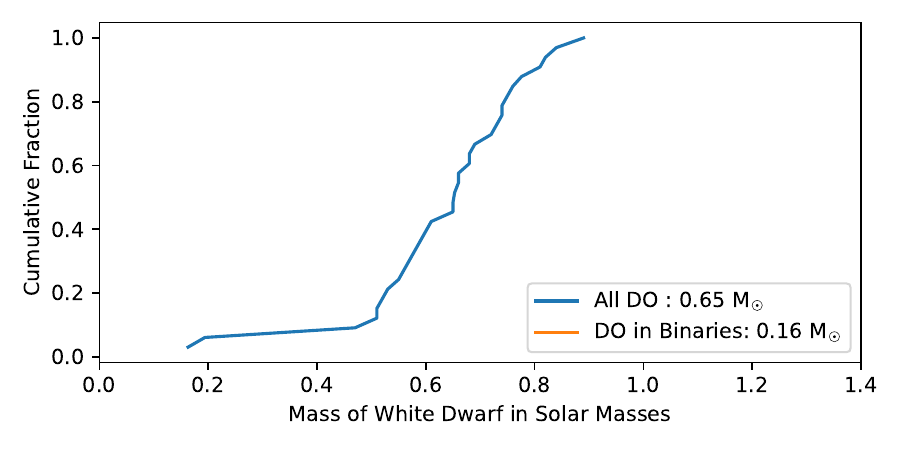}
    \includegraphics[width=\linewidth,clip,trim=0 0 0 0.2in]{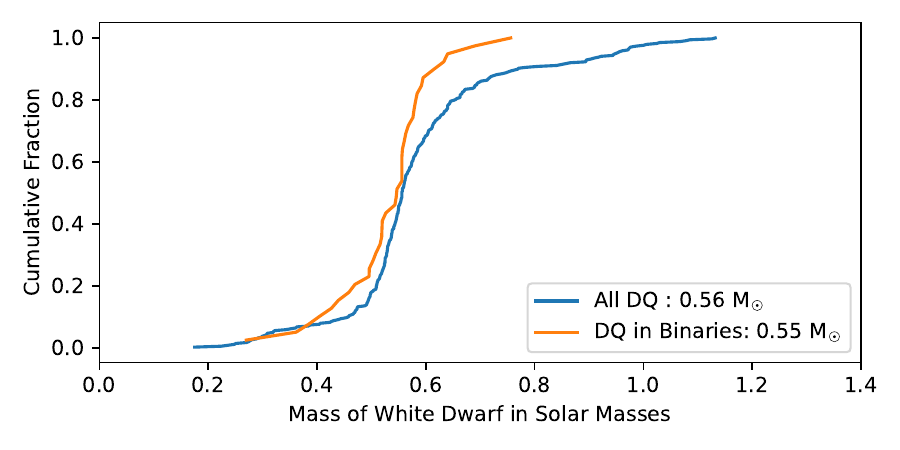}
    \caption{The Cumulative Distribution of Masses of White Dwarfs of Various Spectral Types Obtained by Spectroscopic Modelling. The median mass of each population is presented in the legend. Only a single DO was found in a binary, so no cumulative distribution is depicted.}
    \label{fig:mass_spec}
\end{figure}

\begin{figure}
    \centering
    \includegraphics[width=\linewidth]{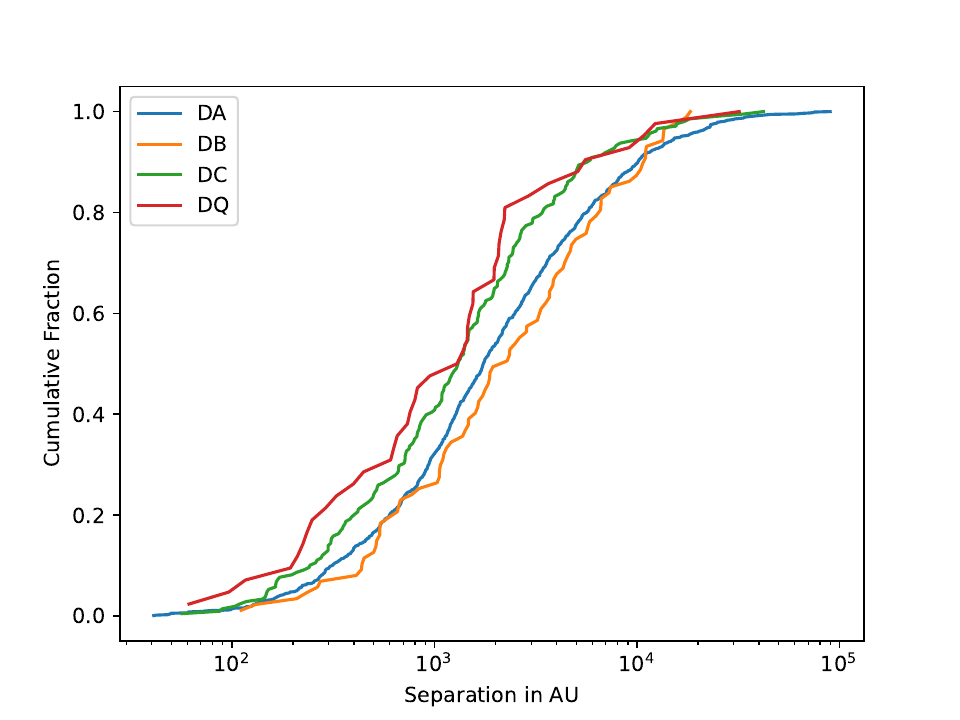}
    \includegraphics[width=\linewidth]{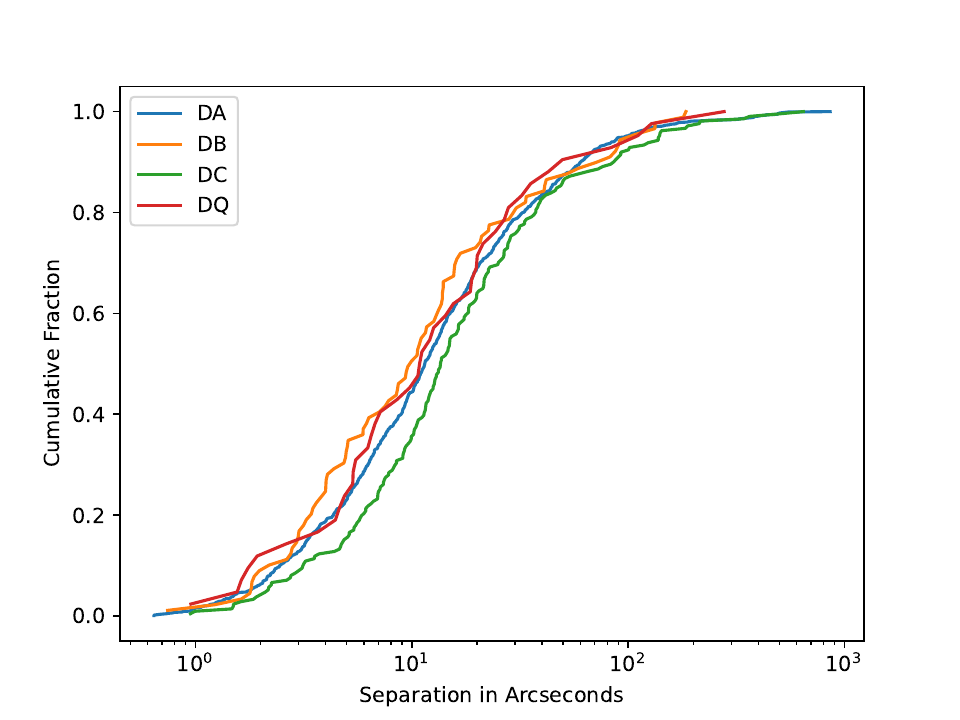}
    \caption{The Cumulative Distribution of Projected Binary Separations of White Dwarfs of Various Spectral Types.  Upper: in physical distance, Lower: angular sepration.}
    \label{fig:sep_spec}
\end{figure}

An examination of the distribution of the masses of the white dwarfs in each class compared to the distribution on those within binaries (Fig.~\ref{fig:mass_spec}) does not uncover potential systematic effects.   The typical masses of DB, DC and DO stars are larger than those of the DA and DQ stars; furthermore, the typical masses of the DA, DB and DQ stars in binaries are similar to the entire population.  The typical masses of DC stars in binaries are slightly less than the general population of DC stars. Both the distribution of DA and DQ stars in binaries are slightly more peaked than the general population.  The masses of DB stars in binaries are consistent with being drawn from the general population.
Focusing on those white-dwarf stars with companions, Figure~\ref{fig:sep_spec} depicts the distribution of the projected separations of the white dwarfs from their companions.  The DA stars have a median separation from their companions of about 1700~AU while the DBs are typically a bit larger at 2300~AU and the DCs and DQs are smaller at 1300~AU.  All of the distributions are approximately log-normal with a width of about a factor of four.   The median angular distance between the white dwarfs and their companions ranges from 10 to 14 arcseconds, sufficiently distant that it would be difficult to argue that the companion affects the spectroscopic classification of the white dwarfs.    In fact the DC white dwarfs which are likely to be the closest to their companions actually appear further in angular separation. 
\section{Conclusions} 
\label{sec:conclus}

The paper examines the simple question of whether the spectral type of a white dwarf correlates with its presence in a binary system.  Surprisingly, the evidence indicates that white dwarfs with hydrogen features in their spectra (DA white dwarfs) are nearly 25\% less likely (at 6.3\%) to have a companion than those white dwarfs that lack hydrogen features (DB, DC, DO and DQ white dwarfs, at 8.0\%).  The subclass of DQ white dwarfs (those that exhibit carbon features) are the most likely (among these classes) to have companions at 10\%.  So what does binarity have to do with the formation of white dwarfs that lack hydrogen atmospheres?  Perhaps nothing.  The DB and DC white dwarfs both in binaries and in general typically are more massive that the DA white dwarfs indicating that they likely were born from more massive stars which are more likely to have companions.  The modest increase in typical mass from $0.57\textrm{M}_\odot$ for DAs to $0.62\textrm{M}_\odot$ for DBs and DCs, corresponds to an increase in the mass of the progenitor from $1.0\textrm{M}_\odot$ to $1.6\textrm{M}_\odot$ \citep{Cummings_2018}, and a corresponding increase in the binary fraction, for a similar range of separations, of the progenitors from 7.5\% to 9\% \citep{2017ApJS..230...15M}.  This explanation leaves out the DQ stars which exhibit the largest binary fraction at 10\%, the closest separations and the smallest typical mass at $0.55\textrm{M}_\odot$. In the case of these stars, perhaps the presence of their progenitors in a binary system led to their manifestation as DQ white dwarfs, or perhaps their masses have been underestimated systematically and the DQ stars originate from even more massive stars than the DB and DC stars with correspondingly larger binary fractions. 

\section*{Acknowledgements}

Simon Blouin posed brought up the question of the binary fraction of white dwarfs after Hiba tu Noor's presentation at EuroWD24. JH acknowledges support from the Natural Sciences and Engineering Research Council of Canada (NSERC) through a Discovery Grant.  This work presents results from the European Space Agency space mission \textit{Gaia}, where data are processed by the \textit{Gaia} Data Processing and Analysis Consortium.

\section*{Data Availability}

The catalogue of candidate binaries identified by \citet{2021MNRAS.506.2269E} is available at \url{https://zenodo.org/records/4435257}.  The \citet{2021MNRAS.508.3877G} catalogues is accessible at \url{https://cdsarc.cds.unistra.fr/viz-bin/cat/J/MNRAS/508/3877}. The Montreal White Dwarf Database \citep{2017ASPC..509....3D} can be downloaded from \url{https://www.montrealwhitedwarfdatabase.org/}.  The catalogues created for this work can be downloaded at \url{https://zenodo.org/records/12803292}.



\bibliographystyle{mnras}
\bibliography{main} 
\end{document}